\documentclass[twoside,12pt]{article}
\usepackage{CJK}
\usepackage{indentfirst}
\usepackage{bm}
\usepackage{graphicx}
\usepackage{amsmath}

\usepackage{epstopdf}
\renewcommand{\u}{\underline}
\renewcommand{\b}{\mathbf}

\renewcommand{\i}{\includegraphics}

\begin{document}
\begin{CJK*}{GBK}{song}

\thispagestyle{empty} \vspace*{0.8cm}\hbox
to\textwidth{\vbox{\hfil\huge\sf Chinese Physics B\hfill}\hfill}
\par\noindent\rule[3mm]{\textwidth}{0.2pt}\hspace*{-\textwidth}\noindent
\rule[2.5mm]{\textwidth}{0.2pt}

\begin{center}
\LARGE\bf Disappearance of the Dirac cone in silicene due to the presence of an electric field\footnotemark[1]
\end{center}

\begin{center}
D~A~Rowlands \ and \ Yu-Zhong~Zhang\footnotemark[2]
\end{center}

\begin{center}
\begin{small} \sl Shanghai Key Laboratory of Special Artificial Microstructure Materials and Technology,
School of Physics Science and Engineering, Tongji University, Shanghai 200092, P.R.~China
\end{small}
\end{center}

\begin{center}
\small (Received X XX XXXX; revised manuscript received X XX XXXX)
\end{center}

\vspace*{2mm}

\begin{center}
\begin{minipage}{15.5cm}
\parindent 20pt\small
Using the two-dimensional ionic Hubbard model as a simple basis for describing the electronic structure of silicene in the presence of an electric field induced by the substrate, we use the coherent-potential approximation to calculate the zero-temperature phase diagram and associated spectral function at half filling. We find that any degree of symmetry-breaking induced by the electric field causes the silicene structure to lose its Dirac fermion characteristics, thus providing a simple mechanism for the disappearance of the Dirac cone.
\end{minipage}
\end{center}

\begin{center}
\begin{minipage}{15.5cm}
\begin{minipage}[t]{2.3cm}{\bf Keywords:}\end{minipage}
\begin{minipage}[t]{13.1cm}
silicene, Dirac cone, ionic Hubbard model, coherent-potential approximation
\end{minipage}\par\vglue8pt
{\bf PACC: }
71.10.-w, 71.10.Fd, 71.30.+h
\end{minipage}
\end{center}

\footnotetext[1]{Project supported by National Natural Science Foundation of China (No.~11174219), Program for New Century Excellent Talents in University (No.~NCET-13-0428), Research Fund for the Doctoral Program of Higher Education of China (No.~20110072110044) and the Program for Professor of Special Appointment (Eastern Scholar) at Shanghai Institutions of Higher Learning as well as the Scientific Research Foundation for the Returned Overseas Chinese Scholars, State Education Ministry.}
\footnotetext[2]{Corresponding author. E-mail: yzzhang@tongji.edu.cn}

\section{Introduction}

Silicene is a silicon-based material which has only recently been experimentally-realized~\cite{Kara2012,Chen2013}. Like carbon-based graphene~\cite{CastroNeto2009}, it has a two-dimensional (2D) structure which gives rise to its unusual and potentially important properties. As a silicon-based material, silicene has the highly-desirable characteristic of being compatible with existing semiconductor technologies~\cite{Wang2014,Cheng2013}, but unfortunately has not yet been observed to exist spontaneously and remains difficult to synthesize. A continuous film or sheet of silicene is of particular interest and has recently been grown on a Ag(111) surface~\cite{Lalmi2010,Lin2012,Chen2012,Vogt2012}. Scanning tunneling microscopy (STM) experiments revealed a 2D graphene-like honeycomb structure but with a very small buckling in the direction of the plane, as had been predicted by earlier density-functional theory-based calculations~\cite{Cahangirov2009}. This structure can be viewed in terms of two triangular sublattices with a unit cell containing two neighbouring sites displaced alternatively perpendicular to the plane. While the existence of freestanding silicene has yet to be realized, the extent to which the electronic properties of silicene can be influenced by the properties of the substrate are being investigated, for example the application of an external stress achieved by adjusting its lattice parameter~\cite{Fleurence2012}. Such an understanding is crucial if the theoretically-predicted important properties of silicene are to be realized in practice.

One of the most fundamental and important questions is whether silicene possesses graphene-like Dirac fermion properties after being grown on a substrate. Recently Chen~{\it{et al}}~\cite{Chen2012} found experimental evidence from STM and STS (scannning tunneling spectroscopy) measurements for the existence of Dirac fermions in silicene grown on a Ag(111) substrate. On the other hand, Lin~{\it{et al}}~\cite{LinChun-Liang2013} reported that silicene sheet loses its Dirac fermion characteristics because of Ag(111) substrate-induced symmetry breaking due to hybridization between the Si and Ag atoms. This was also demonstrated by STM and STS measurements and density functional theory calculations~\cite{LinChun-Liang2013}.

In order to shed light on these conflicting results, in this paper we investigate the possible role of an electric field that could be induced perpendicular to the Si plane while silicene is grown on a substrate such as Ag(111) as suggested by line-profile STM measurements~\cite{Lalmi2010,Vogt2012}. Due to the buckling of the structure perpendicular to the plane, such an electric field would result in an energy-level offset $\Delta$. We show that $\Delta$ opens up a gap at the Fermi level, splitting the Dirac cone into upper and lower energy bands of parabolic nature near the Dirac K points. While Coulomb correlations can close the gap by screening the effects of $\Delta$, we argue that it is not in general possible to restore the linear Dirac fermion nature of the energy bands near the Dirac points.

In order to demonstrate our argument, we describe the effects of $\Delta$ using the ionic Hubbard model~\cite{Hubbard1981,Fabrizio1999,Zhang2003,Xu2005,Garg2006,Kancharla2007,Hoang2010} on the 2D honeycomb lattice. We examine the system using the coherent-potential approximation (CPA)~\cite{Soven1967}, which is equivalent to the ``scattering correction'' of the Hubbard III approximation~\cite{Hubbard1964}. This theory was in fact the first to describe the Mott metal-insulator transition (MIT) at a finite value of the on-site Coloumb interaction strength $U$~\cite{Hubbard1964}, and improves upon the Hartree-Fock approximation by describing static fluctuations in the potential that an electron sees~\cite{Gyorffy1991}. Its shortcomings notwithstanding~\cite{Gebhard1}, the CPA today remains a valuable tool in the study of strongly-correlated systems since it is able to give a useful approximate description of the MIT while being computationally very quick and simple~\cite{Hoang2010,Le2012,Le2013}.

This paper is organised as follows. In section 2, we introduce our simple model for the system. In section 3 we present our CPA results for the paramagnetic solution and calculate the zero-temperature phase diagram as a function of ionic energy $\Delta$ and on-site Coulomb interaction $U$. We investigate the nature of the phase transitions by examining the spectral function around the Dirac points. We discuss the significance of our results and conclude in section 4.

\section{Formalism}

\vspace*{4mm}
\centerline{\i[scale=0.4]{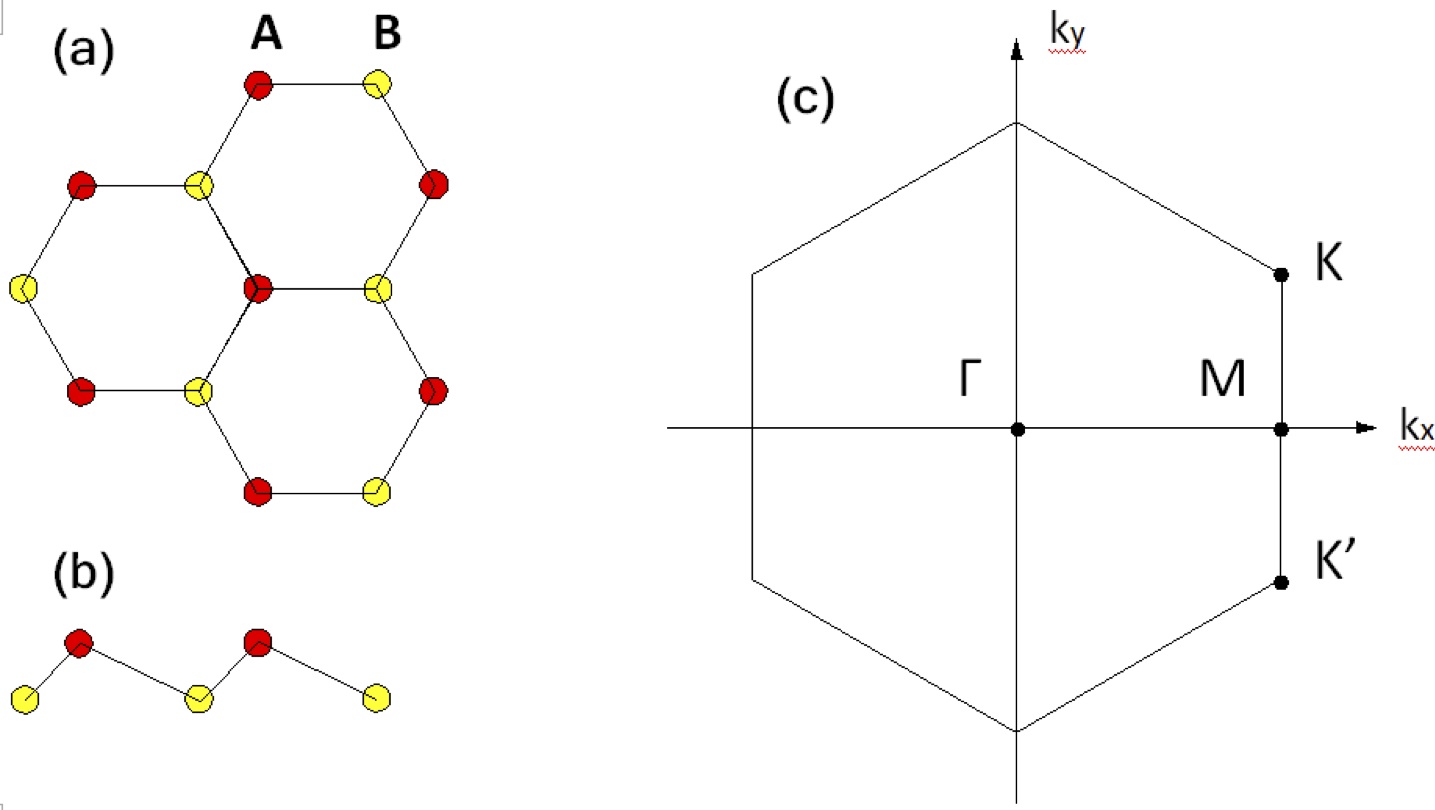}}
\begin{center}
 \parbox{15.5cm}{\small{\bf Fig.1.} (Colour online) Top view (a) and side view (b) of the buckled honeycomb lattice structure with sublattices A and B, and (c) corresponding BZ showing the location of the special points.}
\end{center}

Consider the following ionic Hubbard Hamiltonian for the single-band $\pi$-electrons on a 2D honeycomb lattice with triangular sublattices $A$ and $B$:
\begin{eqnarray}\label{hamiltonian1}
	H = \epsilon_{A}\sum_{i\in{A}}n_{i} + \epsilon_{B}\sum_{i\in{B}}n_{i} + U\sum_{i}n_{i\uparrow}n_{i\downarrow} -t\!\!\! \sum_{i\in{A},j\in{B},\sigma}\!\!\!\left[ c^{+}_{i\sigma} c_{j\sigma}+\textstyle{H.c.}\right]  - \mu\sum_{i}n_{i}
\end{eqnarray}
Here $c^{+}_{i\sigma}\left(c_{i\sigma}\right)$ are the creation (annihilation) operators for an electron with spin $\sigma$ at site $i$, $n_{i\sigma}=c^{+}_{i\sigma}c_{i\sigma}$, and $n_i=n_{i\downarrow}+n_{i\uparrow}$. The nearest-neighbour hopping parameter is denoted by $t$ which we take as the energy unit, the on-site Coulomb repulsion is denoted by $U$, and the chemical potential by $\mu$. The ionic energies are defined by $\epsilon_{A}=\Delta$ and $\epsilon_{B}=-\Delta$ for sublattices $A$ and $B$ respectively.

Such a system has previously been investigated in connection with studies of superconductivity~\cite{Watanabe2013,ZhangLiDa2013}. The buckled honeycomb lattice structure and corresponding Brillouin zone (BZ) are shown in figure 1. For the non-interacting $U=0$ case, the band dispersion is given by
\begin{equation}
	\varepsilon(\b{k})=\pm\sqrt{\Delta^2+t^2\left(3+f(\b{k})\right)}
\end{equation}
with $f(\b{k})=2\cos(\sqrt{3}k_{y}a)+4\cos\left(\frac{\sqrt{3}}{2}k_{y}a\right)\cos\left(\frac{3}{2}k_{x}a\right)$. The effect of finite $\Delta$ is to open up a gap of width $2\Delta$ at the Fermi level and introduce parabolic dispersion character so that the upper $\pi^*$ and lower $\pi$ bands are flattened near the Dirac K points. Switching on the on-site Coulomb interaction $U$ introduces Coloumb correlations described by a self-energy which screens the effect of $\Delta$~\cite{Garg2006}. Nevertheless, we argue that such a self-energy cannot in general cancel the parabolic dispersion character and restore the linear dispersion forming the Dirac cones.

To demonstrate our argument, we apply Hubbard's alloy analogy~\cite{Hubbard1964} to the model. By viewing the system in terms of a disordered alloy for each sublattice where an electron with spin $\sigma$ on a given sublattice sees either a potential $U$ fixed at a site with a spin $-\sigma$ electron present on the same sublattice or zero potential without, the many-body Hamiltonian of equation~(\ref{hamiltonian1}) may be approximated by the one-electron Hamiltonian~\cite{Hoang2010}
\begin{equation}\label{hamiltonian2}
	H = \sum_{i\in{A},\sigma} E_{A\sigma}n_{i\sigma} + \sum_{i\in{B},\sigma}E_{B\sigma}n_{i\sigma} \; -t\!\!\!\sum_{i\in{A},j\in{B},\sigma}\!\!\!\left[ c^{+}_{i\sigma} c_{j\sigma}+\mbox{H.c.}\right]
\end{equation}
where the disorder potential has been defined to include the chemical potential and is given by
\begin{equation}\label{prob}
	E_{\alpha\sigma}=\left\{\begin{array}{l}\epsilon_{\alpha}+{U/2}\;\;\;\mbox{with probability} \;\;\;\langle{n_{\alpha,-\sigma}}\rangle \\  \epsilon_{\alpha}-{U/2}\;\;\;\mbox{with probability}  \;\;\;1-\langle{n_{\alpha,-\sigma}}\rangle \end{array}\right.
\end{equation}
at half-filling with $\mu=U/2$. Here $\langle{n_{\alpha,\sigma}}\rangle$ is the average electron occupancy per site for sublattice $\alpha$ with spin $\sigma$.

In principle, the Green's function corresponding to the Hamiltonian of equation (\ref{hamiltonian2}) needs to be averaged over all possible disorder configurations. This would define an effective medium describing exactly the average properties of a single electron in terms of an exact self-energy $\Sigma_{\alpha}(\b{k})$. Since such an exact average is not feasible, the CPA introduces a simplified effective medium corresponding to a local ($\b{k}$-independent) complex and energy-dependent single-site self-energy or coherent potential $\Sigma_{\alpha}$ which replaces $E_{\alpha}$ on each sublattice $\alpha=A,B$ in the Hamiltonian~(\ref{hamiltonian2}). The medium can be determined by mapping to a coupled single-site impurity problem on each sublattice. The details of the CPA method are given in the appendix.

\section{Results}

First we examine the situation for general $\Delta>0$. For the non-interacting case, the electric field induces a band gap of width $2\Delta$ at the Fermi level and thus the system is a band insulator. This is shown in the example DOS plot of figure 2(d) for $\Delta=0.1$. The parabolic dispersion character of the upper $\pi^*$ and lower $\pi$ bands around the K points are evident from the band structure plot shown in figure 3(d) for the same example $\Delta=0.1$. In order to address the important question as to what happens to the bands as $U$ is increased from zero, we calculated the spectral function near the Dirac K-points using the CPA. For a given point in $\b{k}$-space, the spectral function gives a measure of the disorder-induced broadening and will peak at the band energy eigenvalues. This can be seen on a surface plot along some direction in $\b{k}$-space.

As $U$ is increased, Coulomb correlations screen the energy level offset and the gap gradually closes for all $\Delta>0$. For the example value $\Delta=0.1$, figures 3(d)-(f) and 4(d)-(f) show surface plots along the direction M-K-$\Gamma$ demonstrating how the parabolic character remains as the gap closes until the upper and lower bands are just touching at $U_{c1}\approx{1.54}$.  At exactly the same value, the DOS calculated accurately at the Fermi level $\rho(0)$ becomes finite indicating a transition to a full metallic phase. This is also evident from the increase in contrast around the K point showing the relative increase in spectral weight consistent with a finite DOS. In order to illustrate this phase transition more clearly, figure 5(b) shows how both the charge gap and DOS at the Fermi level $\rho(0)$ vary as a function of $U$. The dotted vertical line indicates the value $U_{c1}\approx{1.54}$ where the gap closes and $\rho(0)$ becomes finite. Note that the values $\rho(0)$ were obtained by first determining the CPA effective medium and site occupation number probabilities self-consistently using energies with an imaginary part $\delta$=0.01. Using the same number probabilities, $\rho(0)$ was then calculated by determining the medium Green's function self-consistently as a function of $\delta$. Extrapolation via polynomial fitting was used in order to obtain the value with $\delta\rightarrow{0+}$. In the case of a finite gap or a semi-metal, $\rho(0)$ extrapolates to zero. In contrast $\rho(0)$ remains finite in the metallic region. Taking into account the tolerances used in the calculations, $\rho(0)\ge{0.01}$ was considered to be a finite value. This method is illustrated in more detail for a selection of values of $U$ in figure 6.

As $U$ is increased further above $U_{c1}$, we find that the bands do not cross and the topology is trivial. The spectral weight becomes smeared as shown in figure 4(e) for the example value $U=2.0$ and eventually the bands separate when the system undergoes a transition to the Mott insulating phase at $U_{c2}\approx{3.5}$. The white horizontal band at the Fermi level shown in figure 4(f) for the example value $U=5.0$ is the Mott insulating gap.

For all $\Delta>0$ we find similar behaviour to that illustrated above. These findings are clear evidence that the system has lost its Dirac fermion characteristics due to the presence of the energy offset $\Delta$, thus explaining why no semi-metallic phase is present. This behaviour arises from symmetry breaking due to a charge density wave. Indeed the DOS plots of figure 2(d)-(f) show the anti-symmetric nature of the local DOS on the $B$-sublattice for the example value $\Delta=0.1$ indicating that electrons generally prefer to be on sublattice $B$. This also produces extra gaps in the Mott insulating phase at the centre of the bands as seen for the example value $U=5.0$.

Now we examine the special case where $\Delta=0$. For both the non-interacting case and for small $U$ the system is semi-metallic with zero gap and DOS at the Fermi level as shown in figure 2(a). The linear dispersion character of the Dirac cones are evident from the spectral function plots in figure 3(a)-(c). When a critical value $U_{c1}\approx{1.32}$ is reached, the DOS becomes finite according to our criterion $\rho(0)\ge{0.01}$ indicating a crossover to a full metallic phase. This is shown in the DOS figure 2(b) for the example value $U=1.54$, and also in figure 5(a) which illustrates how $\rho(0)$ varies as a function of $U$. Unlike the situation where $\Delta$ is finite, here the curve is very smooth with no abrupt change, indicating that this is a smooth crossover from a semi-metal to a full metal rather than an abrupt transition. Eventually at a second critical value $U_{c2}\approx{3.5}$, the system undergoes a phase transition to a Mott insulator as shown in figures 2(c) and 4(c) for the example value $U=5.0$ where the gaps in the centre of the bands seen in the $\Delta=0.1$ calculations are absent.

Our calculations for $\Delta=0$ are in good agreement with the results of Le~\cite{Le2013} who has previously examined the case $\Delta=0$ using the CPA but in combination with a model DOS. However, as noted by Le~\cite{Le2013}, the full metallic phase has not been observed using other approaches. The reason for this may be that due to the fact that the CPA is a theory which performs a static average over disorder and hence tends to favour the metallic state~\cite{Ekuma2013}. Nevertheless, the presence of an intermediate phase between the semi-metal and Mott insulator such as a spin liquid has been found using both quantum monte carlo (QMC)~\cite{Meng2010} and cellular dynamical mean-field theory (CDMFT) calculations~\cite{He2012}, although the validity of such a result has recently been questioned in the QMC case~\cite{Sorella2012}. Furthermore, the Mott transition value of $U_{c2}\approx{3.5}$ obtained in the CPA is in excellent agreement with QMC~\cite{Sorella1992,Meng2010,Sorella2012} and CDMFT calculations~\cite{Wu2010,Liebsch2011,He2012}. Surprisingly, single-site DMFT calculations yielded values which are much higher than expected~\cite{Tran2009,Jafari2009}.

The full phase diagram of the system calculated using the CPA is shown in figure 7. Observe that the metallic phase shrinks as $\Delta$ increases and the Mott transition phase boundary gradually approaches the line $U=2\Delta$. For very large values of $U$, the intermediate metallic phase remains, indicating that it shrinks to a line rather than to a point. In order to clarify the nature of the phase transitions, figure 8 shows the staggered average occupation number (charge density) nB-nA as a function of $U$ for a selection of $\Delta$ values. A charge density wave is present throughout the whole parameter regime and gradually softens with decreasing $\Delta$ and increasing $U$. The smooth nature of the curves indicate that the phase transitions are continuous. For the case $\Delta=0$, this is in agreement with QMC studies~\cite{Sorella1992,Sorella2012} but contrasts with single-site DMFT calculations~\cite{Tran2009} which indicate a first-order transition, and CDMFT calculations which indicate a second order transition~\cite{Wu2010}.

\section{Conclusions}\label{conclusions}

In order to design new electronic devices using silicene, it is essential to know how its electronic properties are affected by interaction with the substrate on which it is grown. In particular, recent experimental investigations have given conflicting results as to whether the 2D graphene-like Dirac fermion properties are preserved~\cite{Chen2012,LinChun-Liang2013,Arafune2013,Chen2013b}.

In order to investigate this important issue, in this paper we have studied the role played by an electric field that could be induced perpendicular to the Si plane while silicene is grown on a substrate such as Ag(111) as suggested by line-profile STM measurements~\cite{Lalmi2010,Vogt2012}. By describing the effect of the field using the 2D ionic Hubbard model on the honeycomb lattice, we have argued that it is not in general possible to restore the linear Dirac fermion nature of the energy bands near the Dirac points. We have illustrated our argument by applying the CPA to the system. We found that while the calculations yield semi-metallic Dirac fermion behaviour for $\Delta=0$ and small on-site Coulomb interaction $U$, the silicene structure loses its Dirac fermion characteristics for all values $\Delta>0$ as a consequence of symmetry breaking induced by a charge density wave.

Although the CPA has a number of shortcomings due to its static nature, we expect future investigations using more sophisticated dynamical methods will further support our argument. It would also be very useful to have an estimate for the parameters $U$ and $\Delta$ to help guide future experimental investigations. We intend to obtain an estimate for these values from first-principles calculations.

In conclusion, our work supports the findings of Lin~{\it{et al}}~\cite{LinChun-Liang2013} and Arafune~{\it{et al}}~\cite{Arafune2013} that silicene sheet grown on a Ag(111) substrate loses its Dirac fermion characteristics. While Lin~{\it{et al}}~\cite{LinChun-Liang2013} explain the result in terms of substrate-induced symmetry breaking due to hybridisation between Si and Ag atoms, our model provides the alternative and very simple mechanism of symmetry breaking due to a substrate-induced electric field.

\section*{Appendix}

For the Hamiltonian of equation~(\ref{hamiltonian2}) with the coherent potential $\Sigma_{\alpha}$ replacing $E_{\alpha}$ on each sublattice $\alpha=A,B$, the CPA average Green's function can be written in the two-sublattice $\b{k}$-space matrix form
\begin{equation}
	\bar{\u{G}}_{\sigma}(\b{k},\omega)=\left[ \begin{array}{cc} \omega-\Sigma_{A\sigma}(\omega) & -t(\b{k}) \\  -t^{*}(\b{k}) & \omega-\Sigma_{B\sigma}(\omega) \end{array}\right]^{-1},
\end{equation}
where
\begin{equation}
	t(\b{k})=1+\exp(i \sqrt{3} k_y a)+\exp(i \frac{\sqrt{3}}{2} k_y a + i \frac{3}{2} k_x a)
\end{equation}
As originally derived by Wallace~\cite{Wallace1947,CastroNeto2009}, the free-electron energy bands on the 2D honeycomb lattice for nearest-neighbour hopping are given by
\begin{equation}\label{wallace}
	\varepsilon_{\pm}(\b{k})=\pm{t}\sqrt{3+f(\b{k})}
\end{equation}
where
\begin{equation}
	f(\b{k})=2\cos(\sqrt{3}k_{y}a)+4\cos\left(\frac{\sqrt{3}}{2}k_{y}a\right)\cos\left(\frac{3}{2}k_{x}a\right)
\end{equation}
and the plus or minus sign corresponds to the upper ($\pi^*$) or lower ($\pi$) bands respectively. Provided expression~(\ref{wallace}) is used, the inverse Green's function diagonal matrix elements can be written in the form
\begin{equation}
	\bar{G}_{\alpha\sigma}(\b{k},\omega)=\frac{\omega-\Sigma_{\bar{\alpha}\sigma}}{(\omega-\Sigma_{A\sigma})(\omega-\Sigma_{B\sigma})-\varepsilon(\b{k})^{2}}
\end{equation}
for each sublattice $\alpha=A(B)$, $\bar{\alpha}=B(A)$, where $\b{k}$ belongs to the first BZ of the sublattice considered and notation has been adopted such that $\alpha\equiv\alpha\alpha$ for clarity. In real space we have
\begin{equation}\label{cpasc1}
	 \bar{G}_{\alpha\sigma}(\omega)=\frac{1}{\Omega_{BZ}}\int_{\Omega_{BZ}}d{\b{k}}\;\frac{\omega-\Sigma_{\bar{\alpha}\sigma}}{(\omega-\Sigma_{A\sigma})(\omega-\Sigma_{B\sigma})-\varepsilon(\b{k})^{2}}
\end{equation}
where the integral is over the first BZ of the sublattice. To determine the medium, the CPA maps to an effective single-site impurity problem. We begin by defining the cavity Green's function $\cal{G}_{\alpha\sigma}(\omega)$ through the relation
\begin{equation}
	{\cal{G}}_{\alpha\sigma}^{-1}(\omega)=\bar{G}_{\alpha\sigma}^{-1}(\omega)+\Sigma_{\alpha\sigma}(\omega)
\end{equation}
for each sublattice $\alpha$, which describes the medium with the self-energy at some chosen site on each sublattice removed i.e.~a cavity. It is not necessary to consider the off-diagonal terms in the Green's function matrix, $\bar{G}_{AB\sigma}$ and $\bar{G}_{BA\sigma}$, since there are no self-energy terms coupling the sublattices. The cavity on each sublattice can now be filled with some real ``impurity'' configuration with impurity Green's functions
\begin{equation}
  G^{\gamma}_{\alpha\sigma}(\omega)=\left[{\cal{G}}_{\alpha\sigma}^{-1}(\omega)-E^{\gamma}_{\alpha\sigma}\right]^{-1}
\end{equation}
with impurity configurations $E^{\gamma=\pm}_{\alpha\sigma}=\epsilon_{\alpha}{\pm}U/2$ as defined by equation (\ref{prob}). The CPA demands that
\begin{equation}\label{cpasc2}
	\langle{G^{\gamma}_{\alpha\sigma}(\omega)}\rangle=\bar{G}_{\alpha\sigma}(\omega),
\end{equation}
where the average is taken over the impurity configuration probablities defined by equation (\ref{prob}), i.e.
\begin{equation}
	\langle{n_{\alpha,-\sigma}}\rangle G^{\gamma=+}_{\alpha\sigma}(\omega) + \left(1-\langle{n_{\alpha,-\sigma}}\rangle\right) G^{\gamma=-}_{\alpha\sigma}(\omega)  =\bar{G}_{\alpha\sigma}(\omega).
\end{equation}
Equations (\ref{cpasc1}) and (\ref{cpasc2}) thus need to be solved self-consistently. Since electrons generally prefer to be on sublattice $B$ for $\Delta\,>0$ and we have $\langle{n_{A}}\rangle+\langle{n_{B}}\rangle=2$ at half-filling where
\begin{equation}
	 \langle{n_{\alpha}}\rangle=\langle{n_{\alpha\uparrow}}\rangle+\langle{n_{\alpha\downarrow}}\rangle=-\frac{1}{\pi}\int_{-\infty}^{0}\mbox{Im}\left[{\bar{G}_{\alpha\uparrow}+\bar{G}_{\alpha\downarrow}}\right]d\omega,
\end{equation}
it is also necessary to ensure the resulting integrated DOS for each sublattice are both consistent with the average occupation number probabilities used in equation~(\ref{cpasc2}), thus adding an extra layer of self-consistency.
For the paramagnetic solution the Green's function is the same for both spin populations so that $G_{\alpha\uparrow} = G_{\alpha\downarrow}$ and $\langle{n_{\alpha\uparrow}}\rangle = \langle{n_{\alpha\downarrow}}\rangle =  \langle{n_{\alpha}}\rangle/2$.



\newpage

\vspace*{4mm}
 \begin{center}
 \addtolength{\tabcolsep}{-5pt}
 \begin{tabular}{cc}
 \i[scale=0.55]{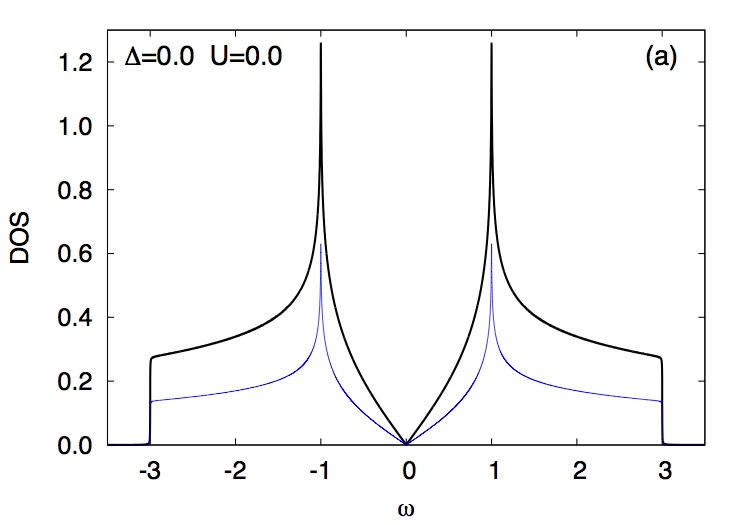} &  \i[scale=0.55]{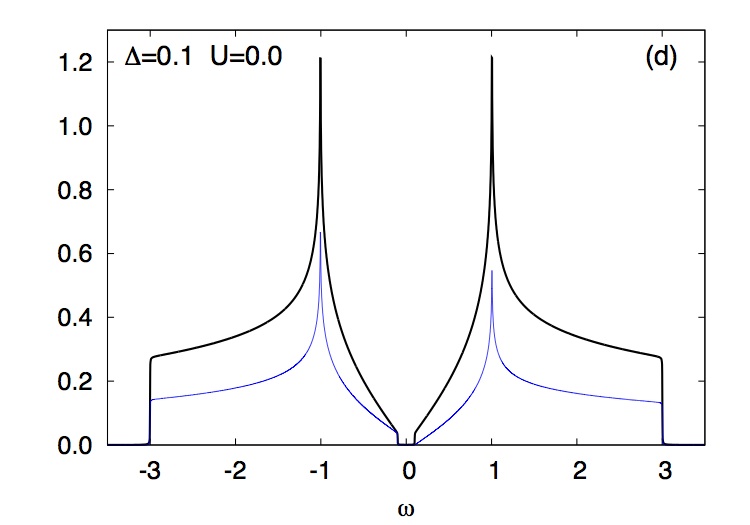} \\
 \i[scale=0.55]{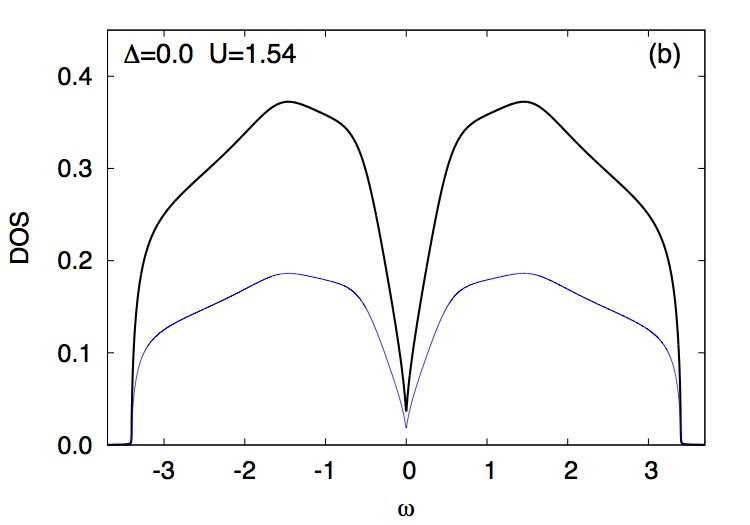} &  \i[scale=0.55]{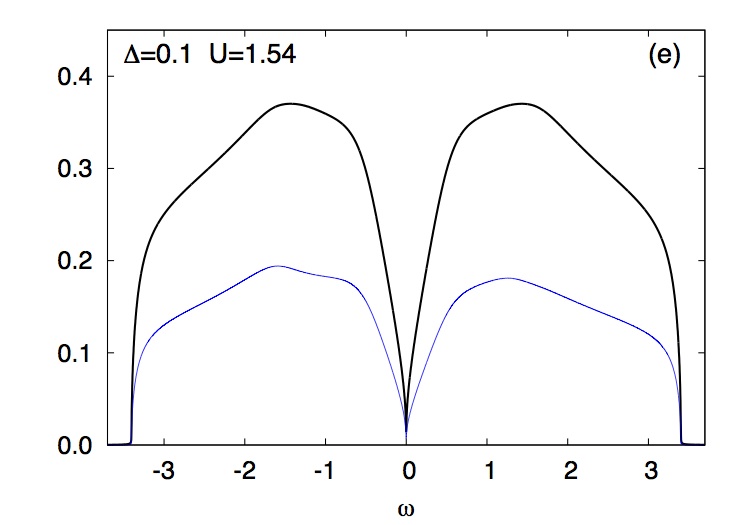} \\
 \i[scale=0.55]{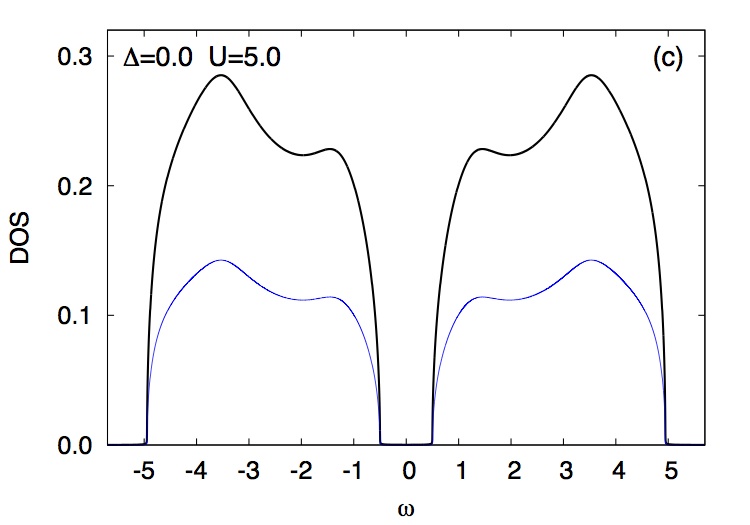} &  \i[scale=0.55]{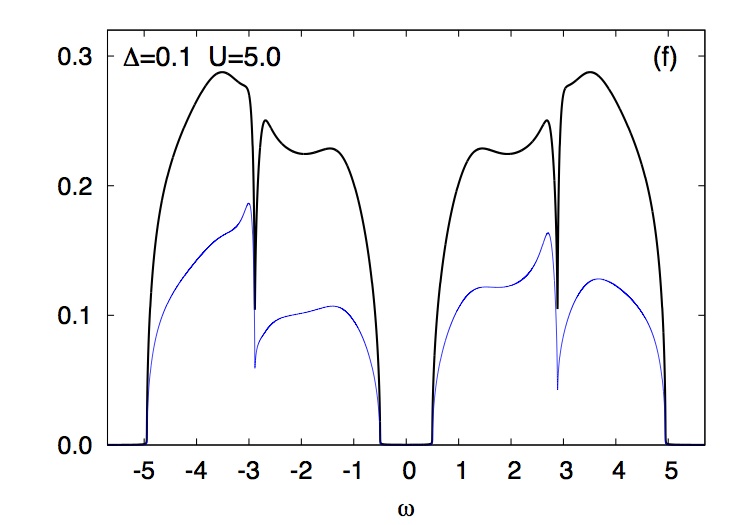}
 \end{tabular}
 \parbox{15.5cm}{\small{\bf Fig.2.} (Colour online) Total DOS per two-site unit cell (black solid line) and local DOS for the $B$ sublattice (blue thin line) for $\Delta=0$ and various values of $U$ for the 2D honeycomb lattice calculated using the CPA. The results shown are for one spin direction only so that integrating the total DOS up to the Fermi level yields one electron. The energy imaginary part (smoothing factor) is $\delta$=0.001. (d)-(f): Same but for $\Delta=0.1$.}
\end{center}
\vspace*{2mm}

\vspace*{4mm}
 \begin{center}
 \addtolength{\tabcolsep}{-10pt}
 \begin{tabular}{cc}
 \i[scale=0.6]{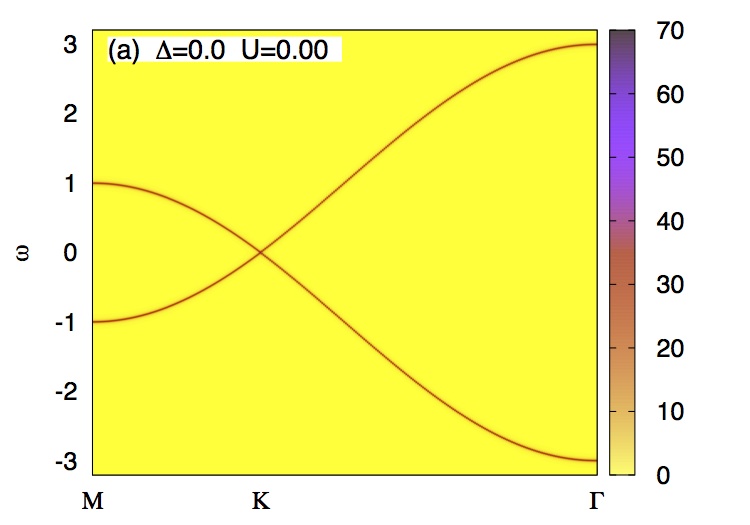} &  \i[scale=0.6]{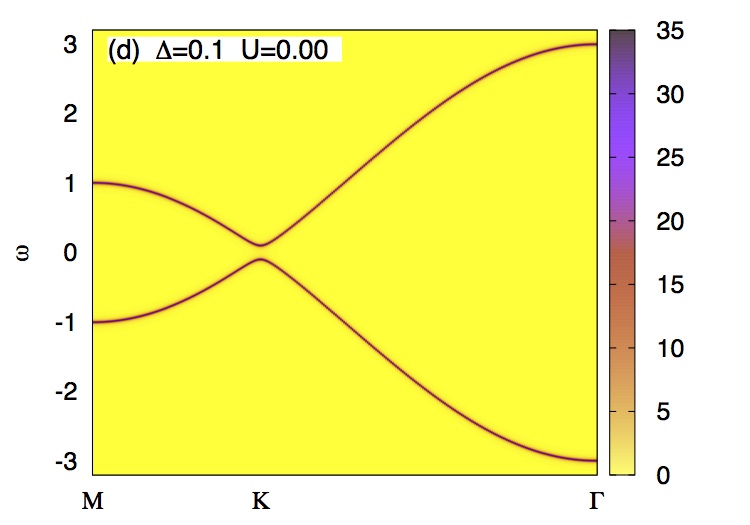} \\
 \i[scale=0.6]{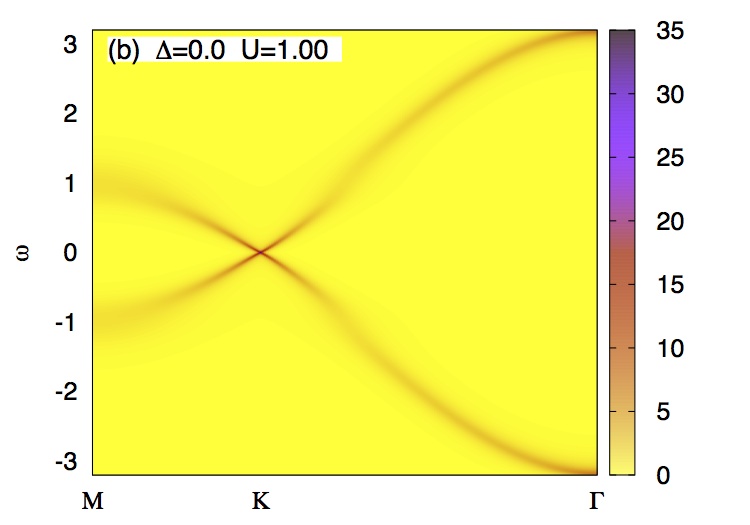} &  \i[scale=0.6]{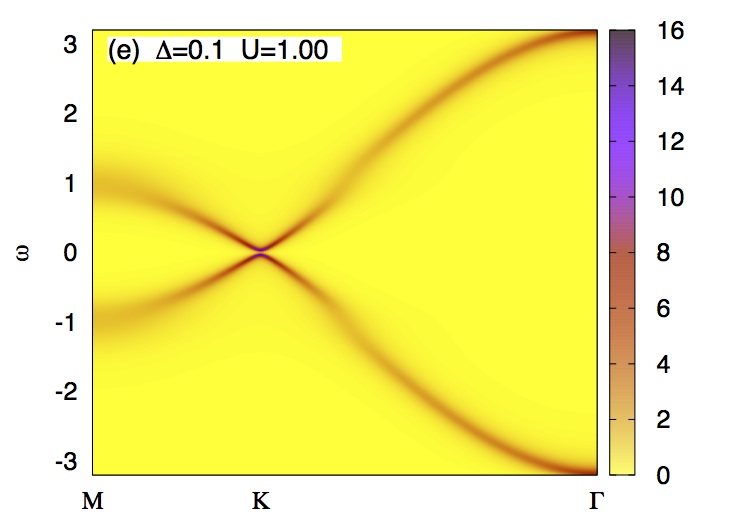} \\
 \i[scale=0.6]{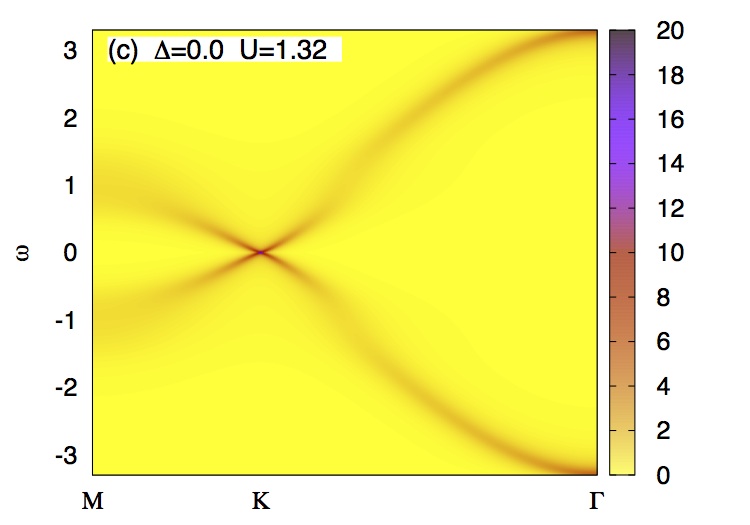} &  \i[scale=0.6]{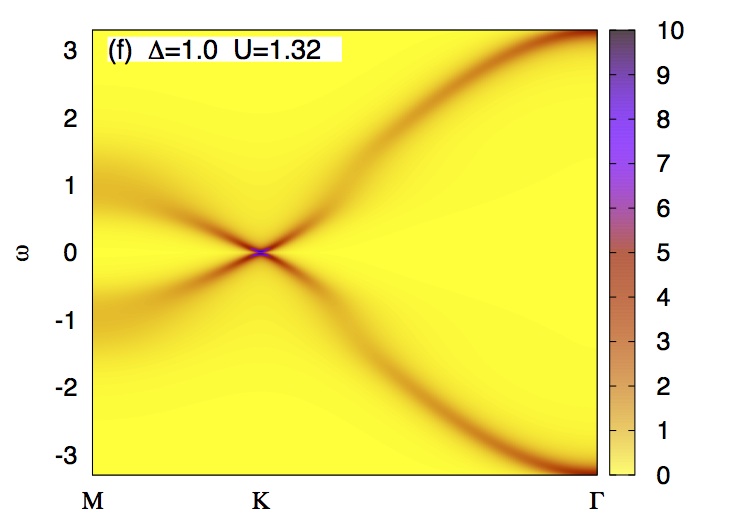}
 \end{tabular}
 \parbox{15.5cm}{\small{\bf Fig.3.} (Colour online) (a)-(c): Spectral function as a function of energy $\omega$ along the direction M-K-$\Gamma$ in $\b{k}$-space for $\Delta=0$ and various values of $U$ for the two-dimensional honeycomb lattice calculated using the CPA. The energy imaginary part (smoothing factor) is $\delta$=0.01. (d)-(f): Same but for $\Delta=0.1$.}
 \end{center}
\vspace*{2mm}

\vspace*{4mm}
 \begin{center}
 \addtolength{\tabcolsep}{-10pt}
 \begin{tabular}{cc}
 \i[scale=0.6]{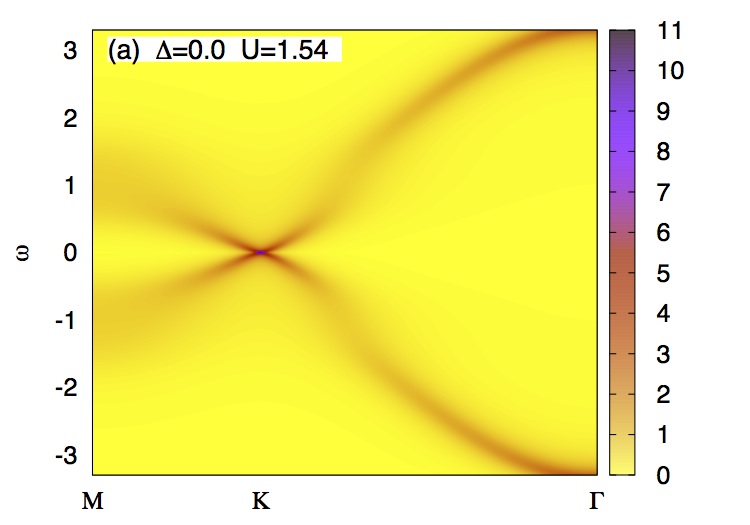} &  \i[scale=0.6]{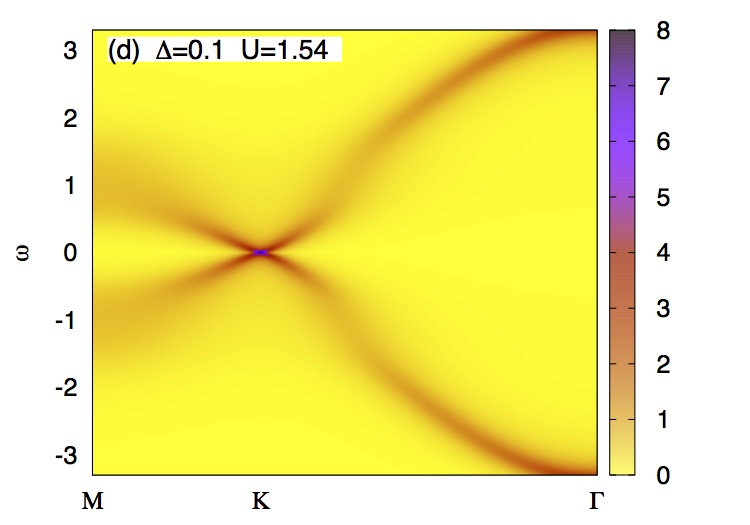} \\
 \i[scale=0.6]{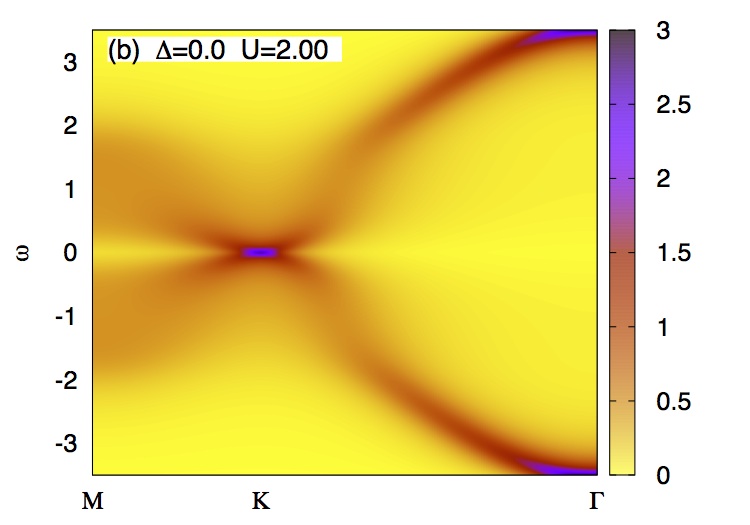} &  \i[scale=0.6]{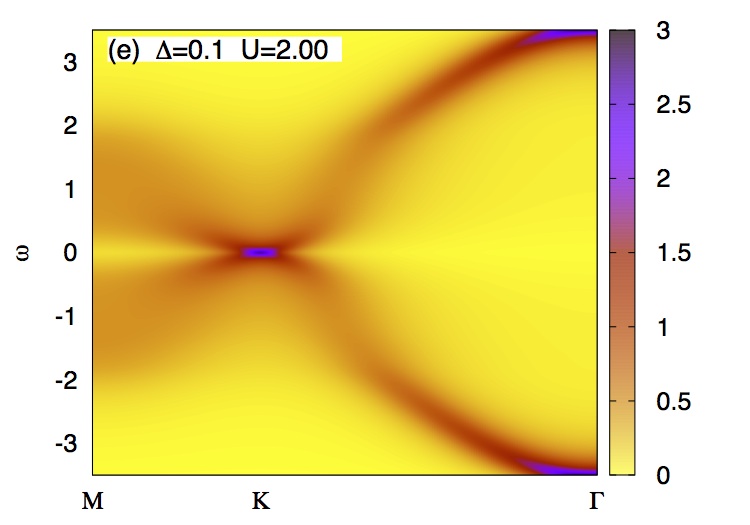} \\
 \i[scale=0.6]{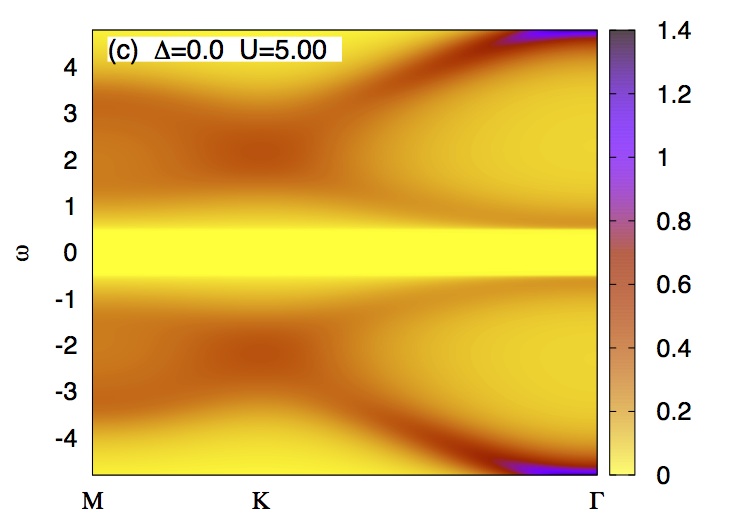} &  \i[scale=0.6]{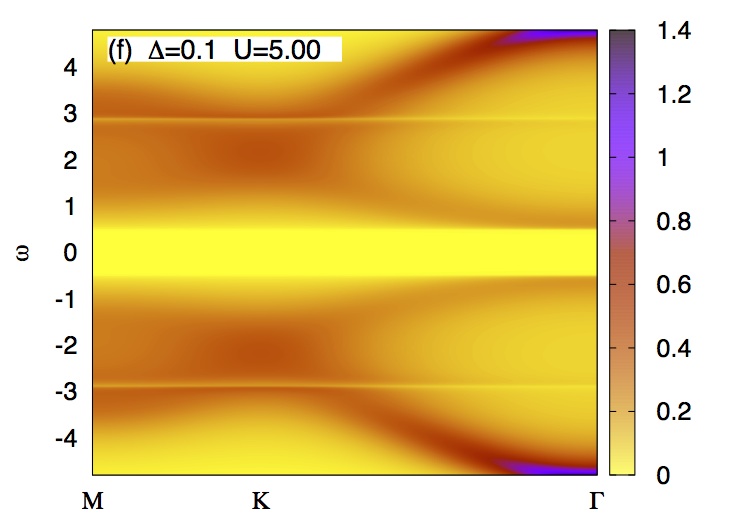}
 \end{tabular}
 \parbox{15.5cm}{\small{\bf Fig.4.} (Colour online) (a)-(c): Spectral function as a function of energy $\omega$ along the direction M-K-$\Gamma$ in $\b{k}$-space for $\Delta=0$ and various values of $U$ for the two-dimensional honeycomb lattice calculated using the CPA. The energy imaginary part (smoothing factor) is $\delta$=0.01. (d)-(f): Same but for $\Delta=0.1$.}
 \end{center}
\vspace*{2mm}

\vspace*{4mm}
 \begin{center}
 \begin{tabular}{c}
 \i[scale=0.75]{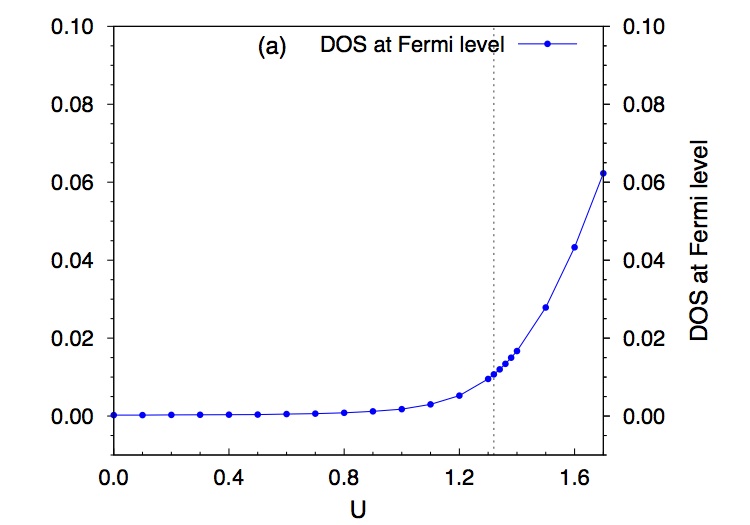} \\ \i[scale=0.75]{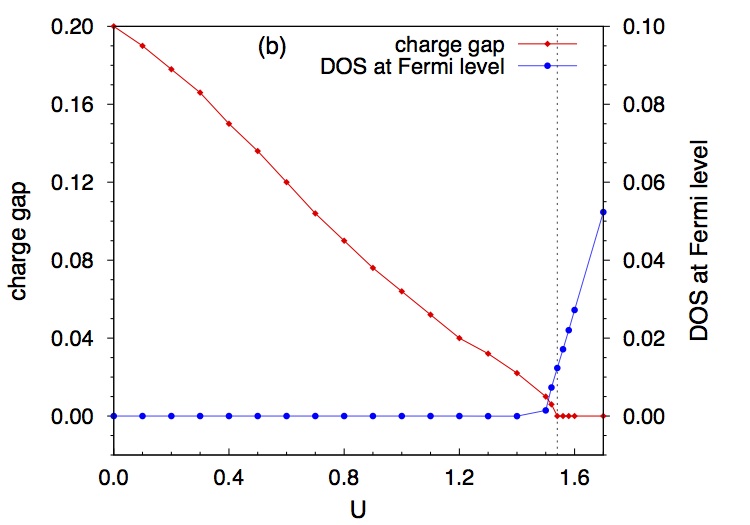}
 \end{tabular}
 \parbox{15.5cm}{\small{\bf Fig.5.} (Colour online) (a) DOS at Fermi level as a function of $U$ calculated for $\Delta=0$ using the CPA. The black vertical dashed line indicates the crossover from semi-metal to metal. (b) Charge gap and DOS at Fermi level as a function of $U$ calculated for $\Delta=0.1$ using the CPA. The black vertical dashed line indicates the band insulator to metal phase transition.}
 \end{center}
\vspace*{2mm}

\vspace*{4mm}
 \begin{center}
 \begin{tabular}{c}
 \i[scale=0.75]{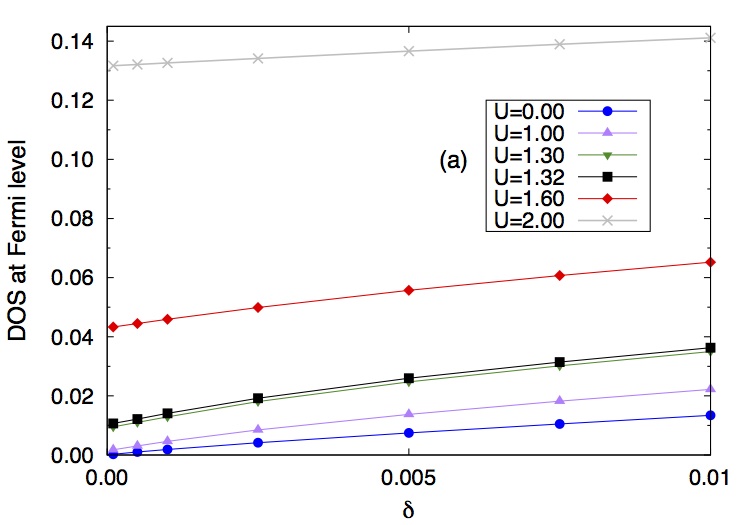} \\ \i[scale=0.75]{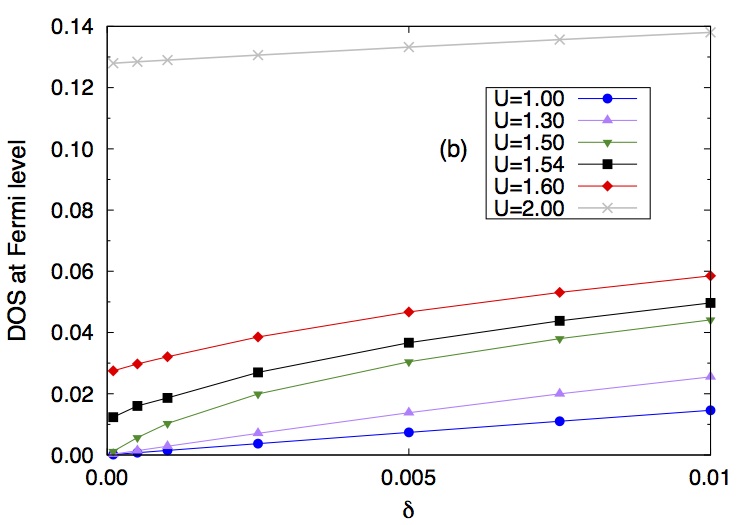}
 \end{tabular}
 \parbox{15.5cm}{\small{\bf Fig.6.} (Colour online) (a) DOS at the Fermi level $\rho(0)$ as a function of imaginary energy part $\delta$ for ionic energy $\Delta=0$ and a selection of on-site interaction values $U$. The curves can be extrapolated to $\delta\rightarrow{0+}$ via polynomial fitting. (b) Same but for $\Delta=0.1$.}
 \end{center}
\vspace*{2mm}

\newpage

\vspace*{4mm}
\centerline{\i[scale=0.8]{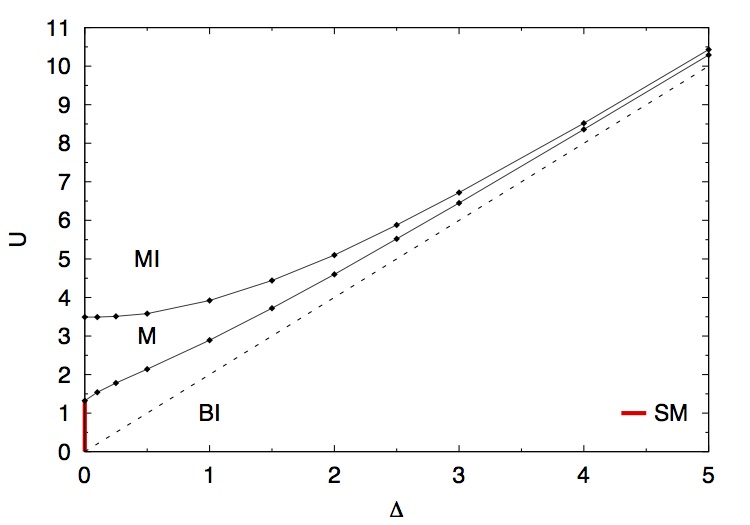}}
\begin{center}
 \parbox{15.5cm}{\small{\bf Fig.7.} (Colour online) $T=0$ phase diagram for the two-dimensional IHM calculated using the CPA. Here SM=semi-metal, BI=band insulator, M=metal, MI=Mott insulator, and the dotted line represents the function $U=2\Delta$.}
\end{center}

\vspace*{4mm}
\centerline{\i[scale=0.8]{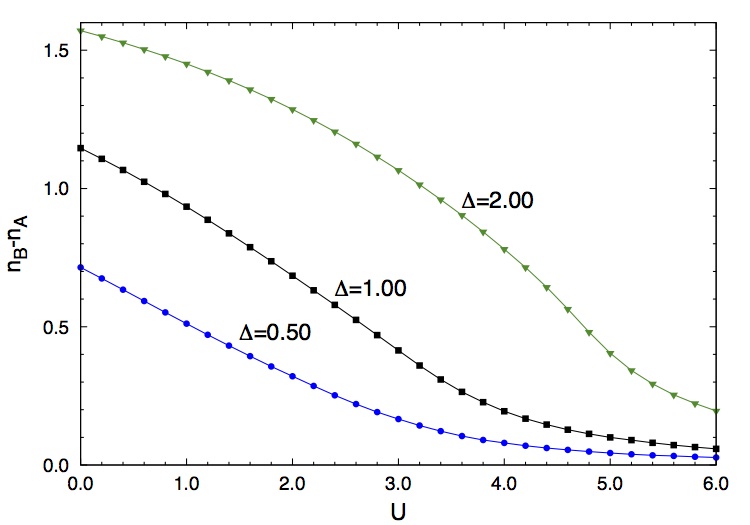}}
\begin{center}
 \parbox{15.5cm}{\small{\bf Fig.8.} (Colour online) Staggered charge density $n_B-n_A$ as a function of $U$ for different values of ionic energy $\Delta$ calculated using the CPA.}
\end{center}

\end{CJK*}

\begin{thebibliography}{99}
\itemsep=-4pt plus.2pt minus.2pt
\small

\bibitem{Kara2012}
Kara A, Enriquez H, Seitsonen A~P, Lew Yan Voon L~C, Vizzini S, Aufray B and Oughaddou H 2012 {\em Surface Science Reports} {\bf 67}  1

\bibitem{Chen2013}
Chen L and Wu K 2013 {\em Physics} {\bf 42} 604

\bibitem{CastroNeto2009}
Neto A~H~C, Guinea F, Peres N~M~R, Novoselov K~S and Geim A~K 2009 {\em Reviews of Modern Physics} {\bf 81} 109

\bibitem{Wang2014}
Wang S~K, Tian H~Y, Yang Y~H and Wang J 2014 {\em Chin. Phys. B} {\bf 23} 017203

\bibitem{Cheng2013}
Cheng G, Liu P~F and Li Z~T 2013 {\em Chin. Phys. B} {\bf 22} 046201

\bibitem{Lalmi2010}
Lalmi B, Oughaddou H, Enriquez H, Kara A, Vizzini S, Ealet B and Aufray B 2010 {\em Appl. Phys. Lett.} {\bf 97}  223109

\bibitem{Lin2012}
Lin C~L, Arafune R, Kawahara K, Tsukahara N, Minamitani E, Kim Y, Takagi N and Kawai M 2012 {\em Appl. Physics Express} {\bf 5} 045802

\bibitem{Chen2012}
Chen L, Liu C~C, Feng B, He X, Cheng P, Ding Z, Meng S, Yao Y and Wu K 2012 {\em Phys. Rev. Lett.} {\bf 109} 056804

\bibitem{Vogt2012}
Vogt P, Padova P~D, Quaresima C, Avila J, Frantzeskakis E, Asensio M, Resta A, Ealet B and Lay G~L 2012 {\em Phys. Rev. Lett.} {\bf 108} 155501

\bibitem{Cahangirov2009}
Cahangirov S, Topsaka M, Akturk E, Sahin H and Ciraci S 2009 {\em Phys. Rev. Lett.} {\bf 102}  236804

\bibitem{Fleurence2012}
Fleurence A, Friedlein R, Ozaki T, Kawai H, Wang Y and Yamada-Takamura Y 2012 {\em Phys. Rev. Lett.} {\bf 108} 245501

\bibitem{LinChun-Liang2013}
Lin C~L, Arafune R, Kawahara K, Kanno M, Tsukahara N, Minamitani E, Kim Y, Kawai M and Takagi N 2013 {\em Phys. Rev. Lett.} {\bf 110} 076801

\bibitem{Hubbard1981}
Hubbard J and Torrance J~B 1981 {\em Phys. Rev. Lett.} {\bf 47} 1750

\bibitem{Fabrizio1999}
Fabrizio M, Gogolin A~O and Nersesyan A~A 1999 {\em Phys. Rev. Lett.} {\bf 83} 2014

\bibitem{Zhang2003}
Zhang Y~Z, Wu C~Q and Lin H~Q 2003 {\em Phys. Rev. B} {\bf 67} 205109

\bibitem{Xu2005}
Xu J, Wang Z~G, Chen Y~G, Shi Y~L and Chen H 2005 {\em Acta Phys. Sin.} {\bf 54} 307

\bibitem{Garg2006}
Garg A, Krishnamurthy H~R and Randeria M 2006 {\em Phys. Rev. Lett.} {\bf 97} 046403

\bibitem{Kancharla2007}
Kancharla S~S and Dagotto E 2007 {\em Phys. Rev. Lett.} {\bf 98} 016402

\bibitem{Hoang2010}
Hoang A~T 2010 {\em J. Phys.: Condens. Matter} {\bf 22} 095602

\bibitem{Soven1967}
Soven P 1967 {\em Phys. Rev.} {\bf 156} 809

\bibitem{Hubbard1964}
Hubbard J 1964 {\em Proc. Roy. Soc. (London)} {\bf A276} 401

\bibitem{Gyorffy1991}
Gy{\"{o}}rffy B~L, Barbiem A, Staunton J~B, Shelton W~A and Stocks G~M 1991 {\em Physica B: Condensed Matter} {\bf 172} 35

\bibitem{Gebhard1}
Gebhard F 1997 {\em The Mott Metal-Insulator Transition: Models and Methods, Springer Series in Solid-State Sciences} (Berlin: Springer)

\bibitem{Le2012}
Le D~A and Hoang A~T 2012 {\em Proc. Natl. Conf. Theor. Phys.} {\bf 37} 73

\bibitem{Le2013}
Le D~A 2013 {\em Modern Physics Letters B} {\bf 27} 1350046

\bibitem{Watanabe2013}
Watanabe T and Ishihara S 2013 {\em J. Phys. Soc. Jpn.} {\bf 82} 034704

\bibitem{ZhangLiDa2013}
Zhang L~D, Yang F and Yao Y 2013 arXiv:1309.7347

\bibitem{Ekuma2013}
Ekuma C~E, Terletska H, Meng Z~Y, Moreno J, Jarrell M, Mahmoudian S and Dobrosavljevic V 2013 arXiv:1306.5712

\bibitem{Meng2010}
Meng Z~Y, Lang T~C, Wessel S, Assaad F~F and Muramatsu A 2010 {\em Nature} {\bf 464} 847

\bibitem{He2012}
He R~Q and Lu Z~Y 2012 {\em Phys. Rev. B} {\bf 86} 045105

\bibitem{Sorella2012}
Sorella S, Otsuka Y and Yunoki S 2012 {\em Scientific Reports} {\bf 2} 992

\bibitem{Sorella1992}
Sorella S and Tosatti E 1992 {\em Europhys. Lett.} {\bf 19} 699

\bibitem{Wu2010}
Wu W, Chen Y~H, Tao H~S, Tong N~H and Liu W~M 2010 {\em Phys. Rev. B} {\bf 82} 245102

\bibitem{Liebsch2011}
Liebsch A 2011 {\em Phys. Rev. B} {\bf 83} 035113

\bibitem{Tran2009}
Tran M~T and Kuroki K 2009 {\em Phys. Rev. B} {\bf 79} 125125

\bibitem{Jafari2009}
Jafari S~A 2009 {\em Eur. Phys. J. B} {\bf 68} 537

\bibitem{Arafune2013}
Arafune R, Lin C~L, Nagao R, Kawai M and Takagi N 2013 {\em Phys. Rev. Lett.} {\bf 110} 229701

\bibitem{Chen2013b}
Chen L, Liu C~C, Feng B, He X, Cheng P, Ding Z, Meng S, Yao Y and Wu K 2013 {\em Phys. Rev. Lett.} {\bf 110} 229702

\bibitem{Wallace1947}
Wallace P~R 1947 {\em Phys. Rev.} {\bf 71} 622

\end{thebibliography}
\end{document}